\documentclass[referee]{jfm}
\usepackage[dvips]{graphicx}
\usepackage{amsmath,amssymb,amsbsy,epsfig}
\usepackage[a4paper, left=2cm, right=2cm, top=2cm, bottom=2cm]{geometry}
\usepackage{natbib}

\title[Dynamic contact angle on an unsaturated wettable porous substrate]
{Dynamic contact angle of a liquid spreading on an unsaturated
wettable porous substrate}

\author[Y.D. Shikhmurzaev and J.E. Sprittles
]
{Y\ls U\ls L\ls I\ls I\ns D.\ns S\ls H\ls I\ls K\ls H\ls M\ls U\ls
R\ls Z\ls A\ls E\ls V\footnote{E-mail: Y.D.Shikhmurzaev@bham.ac.uk}
\and J\ls A\ls M\ls E\ls S\ns E.\ns S\ls P\ls R\ls I\ls
T\ls T\ls L\ls E\ls S\footnote{E-mail: sprittles@maths.ox.ac.uk}}

\affiliation{School of Mathematics, University of Birmingham,
Birmingham B15 2TT, UK,\newline
Mathematical Institute, University of Oxford, Oxford OX1 3LB, UK.}

\begin{document}

\label{firstpage} \maketitle

\begin{abstract}
The spreading of an incompressible viscous liquid over an isotropic
homogeneous unsaturated porous substrate is considered. It is shown
that, unlike the dynamic wetting of an impermeable solid substrate,
where the dynamic contact angle has to be specified as a boundary
condition in terms of the wetting velocity and other flow
characteristics, the `effective' dynamic contact angle on an
unsaturated porous substrate is completely determined by the
requirement of existence of a solution, i.e.\ the absence of a
nonintegrable singularity in the spreading fluid's pressure at the
`effective' contact line. The obtained velocity dependence of the
`effective' contact angle determines the critical point at which a
transition to a different flow regime takes place, where the fluid
above the substrate stops spreading whereas the wetting front inside
it continues to propagate.
\end{abstract}

\section{Introduction}

The modelling of the spreading of liquids over porous substrates in
the framework of continuum mechanics requires, and is based on, the
separation of scales between the `macroscopic' (or `Darcy-scale')
and `microscopic' (or `pore-scale') processes
\citep{Barenblatt-et-al-1990}. In the continuum approximation, the
description of the spreading phenomenon brings in the notion of an
`effective' smooth penetrable solid substrate, which is how the
actual porous medium is represented, together with the notions of
`effective' contact lines and `effective' contact angles that the
free surface of the pure fluid above and the wetting front inside
the substrate form with it. These notions, being averages in the
sense of mechanics of multiphase media \citep{Whitaker-book:1999},
are fundamentally different from the concepts of `contact line' and
`contact angle' used in the modelling of dynamic wetting on the
actual, as opposed to `effective', impermeable solid surfaces. In
particular, the notion of a `static' (or\ `equilibrium') contact
angle, central to the modelling of dynamic wetting where it is a
measure of wettability of the solid \citep{Ralston-et-al:2008},
becomes meaningless for the Darcy-scale description of the liquid
spreading over a porous medium: as experiments show \citep[e.g.\
][]{Clarke-Blake-porous-2002,Starov-Velarde-review-2003,Markicevic:2010},
if, say, a drop of liquid is deposited onto an unsaturated wettable
porous substrate, the eventual equilibrium state will be the drop
completely imbibed into the porous medium, with no liquid left above
it and hence no `effective' static (equilibrium) contact angle
between the, now non-existent, free surface of the pure liquid and
the effective surface of the substrate.

The main theoretical implication of this absence of a meaningful
effective static (equilibrium) contact angle is that the effective
{\it dynamic\/} contact angle that the free surface of a spreading
liquid forms with a porous substrate on the Darcy-scale cannot be
regarded as essentially a perturbation of the static contact
angle, which is what one invariably finds in all models dealing
with the dynamic wetting of impermeable solid surfaces \citep[see,
][for
reviews]{Dussan-review79,deGennes:1985,Blake-review:2006,dry-facts:2011}.
Therefore, it becomes important to consider the problem from first
principles, without implying a-priori that the case of a porous
substrate can be described by adjusting the concepts borrowed from
the modelling of dynamic wetting of impermeable solid surfaces,
such as the equilibrium contact angle.

In the present paper, we show that, unlike the situation one has in
dynamic wetting, where the `microscopic' dynamic contact angle has
to be specified as an additional boundary condition\footnote{Here we
are not discussing the so-called `apparent' contact angle resulting
from the free-surface bending near the contact line.  This angle is
not part of the mathematical problem formulation; it is an auxiliary
concept introduced in some works and in different ways (see, e.g.,
\cite{Wilson-etal06}) to interpret experimental results when the
measured angle differs from the angle imposed in the problem
formulation.}, for the process of the fluid spreading over a porous
substrate one can find the `effective' dynamic contact angle from
the requirement of the absence of a nonintegrable singularity of the
fluid's pressure at the contact line, i.e.\ essentially from the
requirement that a solution exists. (The flow inside the porous
matrix is treated in the standard way, i.e.\ employing a dynamic
wetting model for the impermeable solid surface to describe the
propagation of menisci in the pores.) The limits confining the
considered regime suggest certain experimentally verifiable
predictions of the model which we briefly discuss in the light of
the available experimental data.

\section{\label{Formulation}Problem formulation}

\begin{figure}
\centerline{\epsfig{file=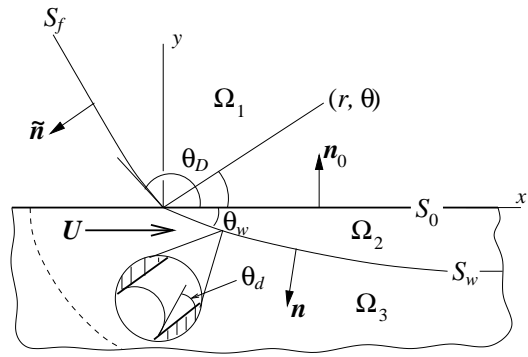,scale=0.38}}
 \caption{Sketch of the spreading of a viscous liquid
 over an unsaturated wettable porous substrate in a coordinate
 frame moving with the contact line in the framework of continuum
 mechanics. The magnified view illustrates the main flow mode on
 the pore scale, where $\theta_d$ is the dynamic contact angle formed
 by the moving meniscus with the wall of the pore.
 Regions $\Omega_1$, $\Omega_2$ and $\Omega_3$
 correspond to the pure liquid, the saturated part of the porous
 medium and the unsaturated porous matrix, respectively;
  $\theta_D$ is the `effective' dynamic contact angle formed by the
  free surface of the liquid $S_f$ with the surface $S_0$ of the
  `effective'  porous substrate, and $S_w$ is the wetting front
  inside the substrate.
   }
\label{sketch}
\end{figure}

Consider an incompressible Newtonian fluid of density $\rho$ and
viscosity $\mu$ spreading at a speed $U$ over an isotropic
homogeneous unsaturated porous substrate characterized by an
effective pore size $a$. The gas displaced by the fluid from the
substrate and from the inside of the porous matrix is assumed to
be ideal and at a constant pressure with respect to which the
pressure in the pure fluid and in the fluid inside the porous
medium will be measured. In order to be able to model the process
in the framework of continuum mechanics, we need a separation of
scales between the macroscopic and the pore-scale phenomena, i.e.\
we have to consider the continuum limit
\begin{equation}
 \label{continuum_limit}
 \epsilon=\frac{a}{L}\to0,
\end{equation}
where $L$ is the characteristic length scale on which the
phenomenon is described. The resulting model will be applicable to
experiments if $a/L\ll1$, with the actual value of $a/L$
determining its accuracy. In the 0th approximation in the above
limit, one has (see Fig.~\ref{sketch}) (a) a macroscopic `wetting
front' as a sharp interface $S_w$ separating the saturated porous
medium $\Omega_2$ from the unsaturated matrix $\Omega_3$, (b) an
effective `contact line' at which the free surface $S_f$,
confining the domain $\Omega_1$ occupied by the pure fluid, and
the wetting front meet, and (c) well-defined `contact angles'
$\theta_D$ and $\theta_w$ that the free surface and the wetting
front form with the `effective' surface $S_0$ of the solid. The
reference frame in which the problem will be considered and the
directions of unit normals $\mathbf{n}$, $\tilde{\mathbf{n}}$ and
$\mathbf{n}_0$ to, respectively, $S_w$, $S_f$ and $S_0$ are shown
in the figure. Importantly, in the scheme outlined above we have
already made an assumption that the {\it two\/} contact lines,
i.e.\ the contact line CL1 formed with the substrate by the free
surface $S_f$ and the contact line CL2 formed with it by the
wetting front $S_w$, coincide.  This is always what happens when
the fluid is first brought in contact with the substrate, and we
will examine what follows from this initial situation and, later,
the conditions when the assumption that CL1 and CL2 coincide no
longer holds.

We will consider the case of small Reynolds and capillary numbers
for the flow of the pure fluid, i.e.\ the limit $Re=\rho L
U/\mu\to0$ and
\begin{equation}
 \label{small_Ca}
 Ca=\frac{\mu U}{\sigma}\to0,
\end{equation}
where $\sigma$ is the surface tension of the fluid-gas interface.
In this limit, to leading order, inertial effects can be
neglected, and, as in the case of the dynamic wetting of an
impermeable solid surface
\citep{Huh-Scriven71,Shikhmurzaev-IJMF:1993}, from the
normal-stress boundary condition on the free surface we have that
near the contact line the free surface ($S_f$) is locally planar,
so that locally $\Omega_1$ is a wedge region. For simplicity we
will also neglect gravity, though its inclusion would not change
the main results of the analysis below.

Scaling the pressure and velocity in $\Omega_1$ with $\mu U/L$ and
$U$ respectively, one has that in $\Omega_1$ the dimensionless
pressure $\tilde{p}$ and velocity $\tilde{\mathbf{u}}$ obey the
Stokes equations
\begin{equation}
 \label{Stokes}
 \nabla\cdot\tilde{\mathbf{u}}=0,
 \qquad
 \nabla\tilde{p}=\nabla^2\tilde{\mathbf{u}},
 \qquad
 (\mathbf{r}\in\Omega_1),
\end{equation}
and on the free surface satisfy the standard kinematic and
tangential-stress boundary conditions,
\begin{equation}
 \label{kinem_tang-stress}
 \tilde{\mathbf{u}}\cdot\tilde{\mathbf{n}}=0,
 \qquad
 \tilde{\mathbf{n}}
 \cdot
 [\nabla\tilde{\mathbf{u}}+(\nabla\tilde{\mathbf{u}})^T]
 \cdot
 (\mathbf{I}-\tilde{\mathbf{n}}\tilde{\mathbf{n}})
 =0,
 \qquad(\mathbf{r}\in S_f),
\end{equation}
%
where $\mathbf{I}$ is the metric tensor.

In the porous medium, the flow is driven by the capillary pressure
in the menisci that collectively form the wetting front, and, for
the problem in question, the characteristic pressure and velocity
are $2\sigma/a$ and $U$, respectively. Using the notation $p$ and
$\mathbf{u}$ for the dimensionless pressure and velocity in
$\Omega_2$, in a frame moving with the contact line we have an
equation of motion in the form of Darcy's law,
\begin{equation}
\label{Darcy_dimless}
 \mathbf{u}-\hat{\mathbf{U}} =-K\nabla p,
 \qquad
 (\mathbf{r}\in\Omega_2),
\end{equation}
where $K=2\sigma\kappa/(\mu aLU)$ is the nondimensionalized
permittivity of the porous matrix ($\kappa$ is the actual
permittivity) and $\hat{\mathbf{U}}=\mathbf{U}/U$ is a unit vector
directed along the velocity of the porous substrate. Given that in
porous media $\kappa\propto a^2$ \citep{Probstein:1989}, we have
that $K=O(\epsilon/Ca)$ and, for the problem to be nontrivial, it
is assumed to be finite in the limits (\ref{continuum_limit}),
(\ref{small_Ca}). In the present context, it is convenient to
define $L$ by setting $K=1$.

Since the porosity $\phi$ in a homogeneous  matrix is constant,
the mass balance equation has the standard form
$\nabla\cdot\mathbf{u}=0$, so that, after substituting
(\ref{Darcy_dimless}) into it, one arrives at Laplace's equation
for $p$ in $\Omega_2$:
\begin{equation}
 \label{Laplace-p}
 \nabla^2p=0,\qquad(\mathbf{r}\in\Omega_2).
\end{equation}

On the wetting front, one has the kinematic condition that the
front propagates with the velocity of the fluid, i.e.\ in the
reference frame described above where the process is steady,
\begin{equation}
 \label{kinem-front}
 \mathbf{u}\cdot\mathbf{n}=0,
 \qquad(\mathbf{r}\in S_w),
\end{equation}
and needs to specify the dynamic condition on the pressure $p$. To
formulate this condition in a general case, one has to consider
the modes of motion the menisci go through on the pore scale as
the macroscopic wetting front propagates
\citep{SS-11a,anomalous-12}. Here we will be considering the
simplest case involving the main, `wetting', mode. In this case,
for an `effective' pore with a circular cross-section the
dimensionless pressure at the front is given by
\begin{equation}
 \label{p-front}
 p=-\cos\theta_d,
\end{equation}
where $\theta_d$ is the contact angle formed on the pore scale by
the representative meniscus with the pore wall (Fig.~\ref{sketch}).
Unlike the simplified approach pioneered by \cite{Washburn-1921},
where it is assumed that $\theta_d$ is constant and equal to the
prescribed static contact angle $\theta_s$, we need to take into
account that, as demonstrated by numerous experiment \citep[e.g.\
see Ch.~3 of][for a review]{TheBook}, $\theta_d$ depends on the
wetting speed, i.e.\ to consider the dependence
\begin{equation}
 \label{theta_d-general}
 (\mathbf{u}-\hat{\mathbf{U}})\cdot\mathbf{n}=U^*_{cl}f(\theta_d),
 \qquad
 (\mathbf{r}\in S_w),
\end{equation}
where $U^*_{cl}$ is the appropriate velocity scale depending on
the material parameters of the system (as all velocities above, we
have it nondimensionalized using $U$). The function $f(\theta_d)$
has to be determined theoretically or empirically. For example,
one can apply $f(\theta_d)$ derived using the theory of dynamic
wetting as a process of interface formation \citep{TheBook}, which
has been shown to reliably describe experimental data, though, in
the present context, any function $f(\theta_d)$ representing the
experimentally observed dependence of the form
(\ref{theta_d-general}) could be used, such as, for instance, the
one that comes from the molecular-kinetic theory of wetting
\citep{Blake-Haynes-1969}. For more information about the dynamic
wetting modelling we refer the reader to a recent review
\citep{dry-facts:2011}.

From the theory of flows with forming interfaces one has
\begin{equation}
\label{f_from_IFM}
 f(\theta_d)=
 \left(
 \frac{
 (1+(1-\rho^s_{1e})\cos\theta_s)(\cos\theta_s-\cos\theta_d)^2}
 {4(\cos\theta_s+B)(\cos\theta_d+B)}
 \right)^{1/2},
 \qquad
 U^*_{cl}\equiv\frac{U_{cl}}{U}
 =\frac{1}{U}
 \left(\frac{\gamma\rho^s_{(0)}(1+4\alpha\beta)}{\tau\beta}\right)^{1/2}
\end{equation}
where $B=(1-\rho^s_{1e})^{-1}(1+\rho^s_{1e}u_0(\theta_d))$,
$$
 u_0(\theta_d)=\frac{\sin\theta_d-\theta_d\cos\theta_d}
 {\sin\theta_d\cos\theta_d-\theta_d},
$$
and $\rho^s_{(0)}$, $\rho^s_{1e}$, $\alpha$, $\beta$, $\gamma$,
$\tau$ are material constants characterizing the contacting media.
Their values for some systems can be found elsewhere
\citep{TheBook,Blake-me02}.

So far, the flow in the pure fluid and in the porous medium were
considered separately, and to link them one has to specify three
boundary conditions at $S_0$. One condition that we obviously have
on this surface is the continuity of mass flux:
\begin{equation}
 \label{continuity-mass-flux}
 (\mathbf{\tilde{u}}-\phi\mathbf{u})\cdot\mathbf{n}_0=0,
 \qquad
 (\mathbf{r}\in S_0).
\end{equation}

For the velocity components parallel to $S_0$, i.e.\ for
$\tilde{\mathbf{u}}_\parallel
\equiv\tilde{\mathbf{u}}\cdot(\mathbf{I}-\mathbf{n}_0\mathbf{n}_0)$
and $\mathbf{u}_\parallel
\equiv\mathbf{u}\cdot(\mathbf{I}-\mathbf{n}_0\mathbf{n}_0)$, a
number of boundary conditions have been discussed in the
literature \citep[e.g.][]{Saffman-1971,Jones-1973,
Murdoch-Soliman-1999, Nield-2009,Auriault-2010}, following the
experiments reported by \cite{Beavers-Joseph-1967} and an
empirical condition these authors proposed. As noted, for example,
by \cite{Auriault-2010}, these conditions are aiming at capturing
the effects of order $O(\epsilon)$, i.e.\ go beyond the classical,
i.e.\ 0th-order, approximation in the continuum limit
(\ref{continuum_limit}). In the 0th approximation, all these
conditions reduce to no-slip for the pure fluid,
\begin{equation}
 \label{no-slip}
 \tilde{\mathbf{u}}_\parallel=\hat{\mathbf{U}}_\parallel,
 \qquad
 (\mathbf{r}\in S_0),
\end{equation}
and it is this condition that we will be using here.

The condition of continuity of pressure on $S_0$ in the
dimensionless form yields that
$$
 p =\tilde{p}\, \frac{\epsilon\, Ca}{2},
$$
and hence, to leading order in the limits (\ref{continuum_limit})
and (\ref{small_Ca}), one has
\begin{equation}
 \label{p_on_S_0}
 p=0,
 \qquad
 (\mathbf{r}\in S_0).
\end{equation}

Importantly, unlike the case of an impermeable solid substrate,
where one has to specify the dynamic contact angle (as we need to
specify $\theta_d$ on the pore scale), here, for the effective
contact angle $\theta_D$, we will require only that the flow
parameters in the porous medium remain regular at the contact line
and that the solution in the pure fluid exists.

\section{\label{Angle}Dynamic contact angle}

Consider the asymptotic behaviour of the solution to
(\ref{Stokes})--(\ref{p_on_S_0}) in the case of a two-dimensional
flow as the distance to the contact line $r\to0$. In the polar
coordinates $(r,\theta)$ shown in Fig.~\ref{sketch}, to leading
order, for the pressure $p$ in the porous medium one has
$$
 \frac{\partial^2p}{\partial r^2}+\frac{1}{r}\frac{\partial p}{\partial r}
 +\frac{1}{r^2}\frac{\partial^2p}{\partial \theta^2}=0,
 \qquad (-\theta_w<\theta<0),
$$
$$
 p(r,0)=0,
 \qquad
 \frac{\partial p}{\partial\theta}(r,-\theta_w)
 =r\sin\theta_w,
$$
where the last condition follows from (\ref{kinem-front}) and
(\ref{Darcy_dimless}). The separable solution to this problem is
obviously given by
\begin{equation}
 \label{p_porous}
 p=r\tan\theta_w\sin\theta,
\end{equation}
so that, using the dynamic boundary condition (\ref{p-front}), we
obtain that
$$
 \theta_d=\arccos(-r\tan\theta_w\sin\theta_w)\to\frac{\pi}{2},
 \qquad\hbox{as }r\to0.
$$
Then, from (\ref{theta_d-general}), where now
$(\mathbf{u}-\hat{\mathbf{U}})\cdot\mathbf{n}=\sin\theta_w$, we
have an equation determining $\theta_w$:
\begin{equation}
 \label{sin-theta_w=}
 \sin\theta_w= U^*_{cl}f(\pi/2).
\end{equation}

In order to consider the flow in the pure fluid, we introduce a
stream function
\begin{equation}
 \label{intro:stream}
 \tilde{u}_r=\frac{1}{r}\frac{\partial\tilde{\psi}}{\partial\theta},
 \qquad
 \tilde{u}_\theta=-\frac{\partial\tilde{\psi}}{\partial r},
\end{equation}
so that for the leading-order term $\tilde{\psi}_1$ of the
asymptotic expansion of $\tilde{\psi}$ as $r\to0$ we have a
biharmonic equation
\begin{equation}
 \label{psi_biharm}
 \left(
 \frac{\partial^2}{\partial r^2}+\frac{1}{r}\frac{\partial }{\partial r}
 +\frac{1}{r^2}\frac{\partial^2}{\partial\theta^2}
 \right)^2\tilde{\psi}_1=0,
 \qquad (0<r,0<\theta<\theta_D),
\end{equation}
together with conditions (\ref{kinem_tang-stress}), i.e.\
\begin{equation}
 \label{psi-free}
 \tilde{\psi}_1(r,\theta_D)=0,
 \qquad
 \frac{\partial^2\tilde{\psi}_1}{\partial\theta^2}(r,\theta_D)=0,
\end{equation}
on the free surface, and conditions (\ref{continuity-mass-flux})
and (\ref{no-slip}), i.e.\
\begin{equation}
 \label{psi-substrate}
 \tilde{\psi}_1(r,0)=r\phi \tan\theta_w,
 \qquad
 \frac{\partial\tilde{\psi}_1}{\partial\theta}(r,0)=r,
\end{equation}
on the surface of the solid substrate. In writing down the first
of conditions (\ref{psi-substrate}) we made use of
(\ref{p_porous}) and integrated along $S_0$.

The separable solution to the problem
(\ref{psi_biharm})--(\ref{psi-substrate}) has the form
\begin{equation}
 \label{stream:soln}
 \tilde{\psi}_1=r(A_1\sin\theta+A_2\theta\sin\theta
  +A_3\cos\theta+A_4\theta\cos\theta),
\end{equation}
where the constants $A_1,\dots,A_4$ are given by
$$
A_1=-\frac{\theta_D}{\sin\theta_D\cos\theta_D-\theta_D}
 -A_3 \frac{\cos^2\theta_D}{\sin\theta_D\cos\theta_D-\theta_D},
$$
\begin{equation}
 \label{A2=}
 A_2=\frac{\sin^2\theta_D}{\sin\theta_D\cos\theta_D-\theta_D}
 +A_3 \frac{\sin\theta_D\cos\theta_D}
 {\sin\theta_D\cos\theta_D-\theta_D},
\end{equation}
\begin{equation}
 \label{A3=_A4=}
 A_3=\phi \tan\theta_w,
 \qquad
 A_4=A_2\cot\theta_D.
\end{equation}

\begin{figure}
\centerline{\epsfig{file=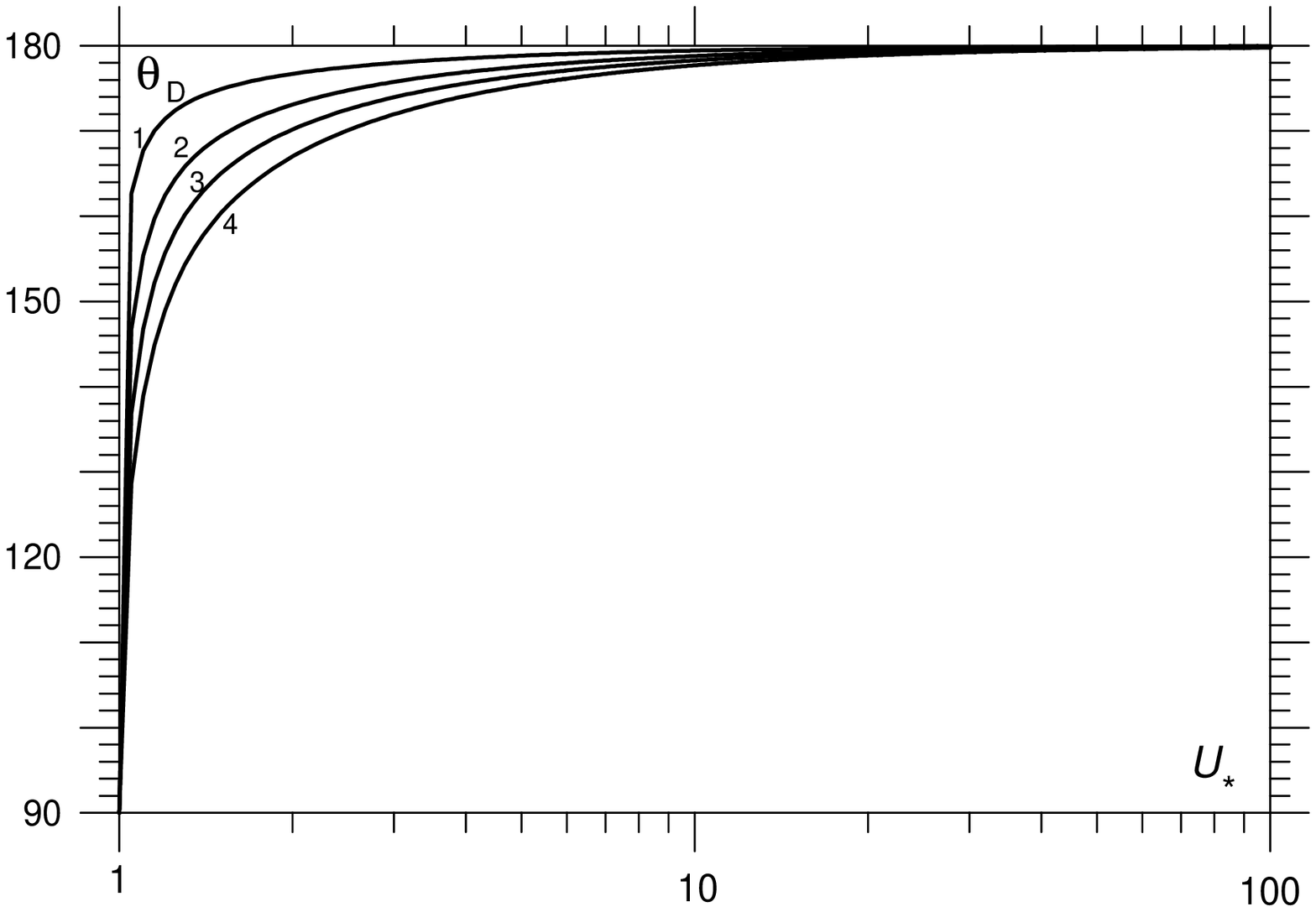,scale=0.5}}
 \caption{Dependence of the effective dynamic contact angle
 $\theta_D$ on the dimensionless contact-line speed $U_*$ for
 different porosities. Curves 1, 2, 3 and 4 correspond to
 $\phi=0.1$, $0.2$, $0.3$ and $0.4$, respectively.
   }
\label{theta_D}
\end{figure}

Using (\ref{intro:stream}), we can write down the radial
projection of the second equation in (\ref{Stokes}) in the form
$$
\frac{\partial\tilde{p}}{\partial r}=
 \left(
 \frac{1}{r}\frac{\partial^3}{\partial r^2\partial\theta}
 +\frac{1}{r^3}\frac{\partial^3}{\partial\theta^3}
 +\frac{1}{r^2}\frac{\partial^2}{\partial r\partial\theta}
 \right)\tilde{\psi}_1,
$$
and, after substituting the solution (\ref{stream:soln}), arrive at
$$
\frac{\partial\tilde{p}}{\partial r}=
-\frac{2}{r^2}(\sin\theta+\cot\theta_D\cos\theta)A_2.
$$
Thus, the leading term in the coordinate expansion of the stream
function will not give rise to a nonintegrable singularity of
pressure if and only if $A_2=0$, or, using (\ref{A2=}) and
(\ref{A3=_A4=}),
$$
 \tan\theta_D=-\phi \tan\theta_w.
$$
Given the expression (\ref{sin-theta_w=}) for $\sin\theta_w$ and
introducing the dimensionless contact-line speed $U_*\equiv
1/(f(\pi/2)U^*_{cl})=U/(f(\pi/2)U_{cl})$, we can also write this
equation down as
\begin{equation}
 \label{theta_D=}
 \theta_D=\pi+\arctan\left( -\frac{\phi}
  {\sqrt{U^2_*-1}}
  \right).
\end{equation}
This equation specifies $\theta_D$ in terms of the speed $U$ of
the contact line with respect to the substrate, the porosity of
the matrix $\phi$ and the material properties of the contacting
media accumulated in $U_{cl}$ and $f(\pi/2)$. The
velocity-dependence of $\theta_D$ for different porosities is
shown in Fig.~\ref{theta_D}. As one can see, this dependence is
much steeper than the velocity-dependence of the dynamic contact
angle for the dynamic wetting of an impermeable substrate
\citep[e.g.\ see Ch.~3 of][for a review]{TheBook}. In particular,
$d\theta_D/dU_*\to+\infty$ as $U_*\to1+$.

The dynamic contact angle given by (\ref{theta_D=}) ensures
integrability of the normal stress on the free surface and hence the
existence of a solution. The next term in the asymptotic expansion
of $\tilde{\psi}$ has the form $\tilde{\psi}_2=r^2F(\theta)$ and
therefore can give rise, at most, to a logarithmic (i.e.\
integrable) singularity of $\tilde{p}$ at the contact line, which
does not affect the existence of the solution.

The obtained result has a clear physical meaning. In the case we
are considering, the dynamics of imbibition determines the normal
to the substrate component of the pure fluid's velocity
independently of the pure fluid's bulk flow, and, given that the
tangential component of the fluid's velocity on the surface of the
substrate is prescribed, as it satisfies the no-slip condition
with a known speed of the substrate, we have a moving contact-line
problem where on the solid surface both components of velocity are
set. In this situation, in the pure fluid the solution exists only
if from the family of stream functions described by
(\ref{stream:soln}) we choose the one that corresponds to a
uniform flow ($A_2=0$, $A_4=0$) with the projections of velocity
on the normal and tangential to the substrate directions equal to
the speed of imbibition and the speed of the solid substrate,
respectively.

The solution (\ref{theta_D=}) for a steady spreading of a fluid over
an unsaturated substrate ceases to exist when the dimensionless
velocity $U_*$ becomes equal to $1$ and both $\theta_D$ and
$\theta_w$ reach $\pi/2$. For $U_*<1$ we have a different regime
with the two contact lines CL1 and CL2 no longer coinciding as the
wetting front moves ahead (dashed line in Fig.~\ref{sketch}), and
the pure fluid finds itself on a saturated substrate. Then, in the
vicinity of CL1 the pressure $p$ satisfies Laplace's equation
$\nabla^2p=0$ together with the boundary conditions
$$
p(r,0)=0, \qquad \frac{\partial p}{\partial\theta}(r,-\pi)=0,
$$
and the local solution with the regular pressure gradient has the
form
\begin{equation}
\label{p_near}
p=
 \sum\limits_{n=1}^\infty C_n r^{1/2+n}\sin\left(\frac{1}{2}+n\right)\theta,
\end{equation}
where $C_n$, $n=1,2,\dots$ are constants. Then, according to
(\ref{Darcy_dimless}) and (\ref{continuity-mass-flux}), one has
$\tilde{u}_\theta(r,0)\propto r^{1/2}\to0$ as $r\to0$. As a result,
if $\tilde{u}_r(r,0)=O(1)$ as $r\to0$, then, to leading order as
$r\to0$, we will have the `classical' moving contact-line problem,
with no imbibition and the no-slip boundary condition on the solid.
This problem, as is well known \citep[see, e.g., Ch.~3
of][]{TheBook}, has no solution. The only way out of this situation
is to conclude that, as CL1 and CL2 separate, the contact line CL1
stops moving. Then, for the pure fluid one simply has a static
contact line with the imbibition velocity $\propto r^{1/2}$ near it
and $\theta_D=\pi/2$. It is easy to verify that the solution to this
problem exists. As CL1 and CL2 separate and CL2 moves ahead, the
imbibition process near CL1 slows down, leading to $C_1=0$ in
(\ref{p_near}) and hence $\tilde{u}_\theta\propto r^{3/2}$, so that
$\theta_D$ no longer has to be equal to $\pi/2$ and can go down as
the imbibition continues.

\section{Discussion}

The described scenario is in agreement with the available
experimental data. In experiments on the spreading drops, it has
been observed that the regime where the drop base and the saturated
area underneath it expand together is followed by the regime where
the two contact lines, CL1 and CL2, separate, as drop's base stops
expanding whereas the saturated area continues to grow
\citep{Clarke-Blake-porous-2002,Starov-Velarde-review-2003,Keshav-Basu-2007,Markicevic:2010}.
This is usually attributed to the `competition' between spreading
and imbibition, and the above analysis shows what this actually
means.

\cite{Markicevic:2010} report that in their experiment the spreading
droplet maintained the shape of a spherical cap throughout the
process and that its base stopped expanding when ``the droplet is a
half of sphere''. In other words, it stopped expanding when the
dynamic contact angle $\theta_D$ became equal to $90^\circ$. This is
exactly what follows from the result of our analysis.

The dependence of the effective contact angle $\theta_D$ on the
contact-line speed for $U_*>1$ given by (\ref{theta_D=}) and
illustrated in Fig.~\ref{theta_D} awaits its experimental
verification. The issue here is that, to extract this dependence
from experiments with unsteady flows, one has to deal with very
short time intervals ($\ll40$~ms according to
\cite{Markicevic:2010}) with a well-controlled spatial resolution
and conditions of the Darcy-scale description satisfied. The problem
is made more complicated by the fact that, as one can see in
Fig.~\ref{theta_D}, the velocity-dependence of $\theta_D$ is very
steep, which brings in additional conditions on the temporal
resolution of the experiments. On the other hand, experiments with
steady flows of the kind commonly used to study dynamic wetting of
impermeable solid surface require sizeable substrates with
well-reproducible properties. To date, no systematic measurements of
$\theta_D$ with controlled spatial resolution have been reported.

As we have shown, the requirement of the absence of a
nonintegrable singularity of pressure in the pure fluid at the
contact line, which is equivalent to the requirement of the
existence of a solution, uniquely specifies the
velocity-dependence of the `effective' dynamic contact angle
$\theta_D$ formed by the free surface and the `effective' surface
of the porous substrate. In practical computations, it may be
convenient to set (\ref{theta_D=}) as a boundary condition while
considering the normal-stress condition on the free surface as the
equation determining the free surface shape. One implication of
the obtained result is that, if one specifies $\theta_D$ as an
additional boundary condition different from (\ref{theta_D=}),
then, for a solution to exist, it will become necessary to use a
slip boundary condition instead of the no-slip condition
(\ref{no-slip}) employed here \citep{Davis-Hocking-porous:2000}.
As mentioned earlier, this would mean bringing in effects of
$O(\epsilon)$ into the essentially classical, i.e.\ $O(1)$ as
$\epsilon\to0$, formulation. If the velocity-dependence of
$\theta_D$ different from (\ref{theta_D=}) is imposed together
with the no-slip condition (\ref{no-slip}) for the pure fluid
\citep{Alleborn-Raszillier:2004,Reis-et-al:2004}, then there will
be no solution to the problem, although this fact can be hidden
behind, and as a result masked by, simplifying assumptions made in
the process of finding the solution (e.g.\ lubrication
approximation) and the numerical implementation.

From the theoretical viewpoint, it would be interesting to consider
the essentially unsteady process of separation of CL1 and CL2, which
marks the transition between the two regimes described above. As
with every finite-time transition, this is a challenging problem
deserving a detailed investigation.

\bibliographystyle{jfm}
\bibliography{references}

\end{document}